\documentclass{elsart}
\usepackage[dvips]{graphicx}
\def\d0b{\overline{D}\!\,^0}

\def\PLB{{Phys. Lett.} B}
\def\PRL{Phys. Rev. Lett.}
\def\PRD{{Phys. Rev.} D}

\def\EPJ{{Eur. Phys. J.} C}

\newcommand{\mysection}[1]{\section{#1}}

\newcounter{saveeqn}%

\begin{document}
\begin{frontmatter}
\title{Charm System Tests of CPT and Lorentz Invariance with FOCUS}
The FOCUS Collaboration%
\footnote{See \textrm{http://www-focus.fnal.gov/authors.html} for
additional author information.}
\author[ucd]{J.~M.~Link}
\author[ucd]{M.~Reyes}
\author[ucd]{P.~M.~Yager}
\author[cbpf]{J.~C.~Anjos}
\author[cbpf]{I.~Bediaga}
\author[cbpf]{C.~G\"obel}
\author[cbpf]{J.~Magnin}
\author[cbpf]{A.~Massafferri}
\author[cbpf]{J.~M.~de~Miranda}
\author[cbpf]{I.~M.~Pepe}
% ...

\address[ucd]{University of California, Davis, CA 95616}
\address[cbpf]{Centro Brasileiro de Pesquisas F\'isicas, Rio de Janeiro, RJ, Brasil}
\address[cinv]{CINVESTAV, 07000 M\'exico City, DF, Mexico}
\address[cu]{University of Colorado, Boulder, CO 80309}
\address[fnal]{Fermi National Accelerator Laboratory, Batavia, IL 60510}
\address[fras]{Laboratori Nazionali di Frascati dell'INFN, Frascati, Italy I-00044}
\address[ui]{University of Illinois, Urbana-Champaign, IL 61801}
\address[iu]{Indiana University, Bloomington, IN 47405}
\address[korea]{Korea University, Seoul, Korea 136-701}
\address[kp]{Kyungpook National University, Taegu, Korea 702-701}
\address[milan]{INFN and University of Milano, Milano, Italy}
\address[nc]{University of North Carolina, Asheville, NC 28804}
\address[pavia]{Dipartimento di Fisica Nucleare e Teorica and INFN, Pavia, Italy}
\address[pr]{University of Puerto Rico, Mayaguez, PR 00681}
\address[sc]{University of South Carolina, Columbia, SC 29208}
\address[ut]{University of Tennessee, Knoxville, TN 37996}
\address[vu]{Vanderbilt University, Nashville, TN 37235}
\address[wisc]{University of Wisconsin, Madison, WI 53706}

\nobreak
\begin{abstract}
We have performed a search for CPT violation in neutral charm meson
oscillations.  
While flavor mixing in the charm sector is predicted to be 
small by the Standard Model, it is still possible to 
investigate CPT violation through a study
of the proper time dependence of a CPT asymmetry in
right-sign decay rates for $D^0\rightarrow K^-\pi^+$ and
$\d0b\rightarrow K^+\pi^-$. This asymmetry is related
to the CPT violating complex parameter $\xi$ and the
mixing parameters $x$ and $y$: $A_{CPT}\propto{\rm Re}\,\xi\,y-{\rm Im}\,\xi\,x$.
Our 95\% confidence level limit
is $-0.0068<{\rm Re}\,\xi\,y-{\rm Im}\,\xi\,x<0.0234$.
Within the framework of the Standard Model Extension incorporating general
CPT violation, we also find 95\% confidence level limits for the
expressions involving coefficients of Lorentz violation of 
$(-2.8<N(x,y,\delta)(\Delta a_0 + 0.6\,\Delta
a_Z)<4.8)\times 10^{-16}$~GeV, $(-7.0<N(x,y,\delta)\Delta a_X<3.8)\times 10^{-16}$~GeV,
and $(-7.0<N(x,y,\delta)\Delta a_Y<3.8)\times 10^{-16}$~GeV, where
$N(x,y,\delta)$ is the factor which incorporates mixing
parameters $x$, $y$ and the doubly Cabibbo suppressed to Cabibbo
favored relative strong phase $\delta$.
\end{abstract}
\end{frontmatter}
\newpage
\newpage

\mysection{Introduction}

The combined symmetry of charge conjugation (C), parity (P), and 
time reversal (T) is believed to be respected
by all local, point-like, Lorentz covariant field 
theories, such as the Standard Model.  
However, extensions to the Standard Model based on string 
theories do not necessarily require CPT invariance,
and observable effects at low-energies may be within
reach of experiments studying flavor 
oscillations~\cite{kostelecky-95,xing}.
A parametrization~\cite{kostelecky} in which 
CPT and T violating parameters
appear has been developed which allows
experimental investigation in many physical systems including 
atomic systems, Penning traps, and neutral meson 
systems~\cite{alanbook}. Using this parameterization we
present the first experimental results for CPT violation in the charm
meson system.

Searches for CPT violation have been
made in the neutral kaon system. Using an earlier CPT
formalism~\cite{lavoura,kostelecky99}, KTeV
reported a bound on the CPT figure of merit
$r_K \equiv |m_{K^0} - m_{\overline{K}\!\,^0}|/m_{K^0} < (4.5 \pm 3) \times
10^{-19}$~\cite{ktev}. A more recent analysis, using
framework~\cite{kostelecky} and more data extracted limits
on the coefficients for Lorentz violation of 
$\Delta a_{X}, \Delta a_{Y}<9.2\times 10^{-22}$~GeV~\cite{ktevlv}.  
CPT tests in $B^0$ meson decay have been 
made by OPAL at LEP~\cite{opal}, and by
Belle at KEK which has recently reported 
$r_B \equiv |m_{B^0} - m_{\overline{B}\!\,^0}|/m_{B^0} < 1.6 \times
10^{-14}$~\cite{belle}.

To date, no experimental search for CPT violation has been
made in the charm quark sector.  This is due in part to 
the expected suppression of $D^0-\d0b$  oscillations
in the Standard Model, and the lack of a strong mixing
signal in the experimental data.  Recent mixing searches include
a study of lifetime differences between charge-parity (CP) eigenstates~ 
\cite{focusycp,belleycp,babary}, a study
of the time evolution of $D^0$ decays by CLEO~\cite{cleo}
and a study of the doubly Cabibbo suppressed ratio ($R_{\rm DCS}$)  for the
decay $D^0\rightarrow K^+\pi^-$~\cite{link}.
Even without knowledge of the mixing parameters,
one can investigate CPT violation through a study of the time dependence
of $D^0$ decays.
The time evolution of neutral-meson state is governed by a $2\times2$
effective Hamiltonian $\Lambda$ in  the Schr\"odinger
equation. Indirect CPT violation occurs if and
only if the difference of diagonal elements of $\Lambda$ is nonzero. 
The complex parameter $\xi$ controls the CPT violation and is defined as
$\xi=\Delta\Lambda/\Delta\lambda$, where $\Delta\Lambda
=\Lambda_{11}-\Lambda_{22}$ and $\Delta\lambda$ is the difference in
the eigenvalues. $\xi$ is phenomenologically introduced and therefore
independent of the model.    
Time dependent decay probabilities into right-sign~($D^0\rightarrow
K^-\pi^+$) and wrong-sign decay modes
(wrong sign is used here in the context of decays via
mixing, $D^0\rightarrow \d0b\rightarrow K^+\pi^-$)
for neutral mesons which express the CPT
violation have been developed in a general 
parametrization~\cite{kostelecky}.
For the decay of a $D^0$ to a right-sign final state $f$, 
the time dependent decay probability is:
\begin{eqnarray}
P_f(t) & \equiv & | \langle f | T | D^0(t) \rangle |^2={1\over{2}}
       |F|^2 {\rm exp}({-{\gamma \over{2}}t})
\nonumber  \\ 
       &\times &[ ( 1 + |\xi|^2 ) {\rm cosh} \Delta\gamma t/2 + 
           (1 - |\xi|^2 ) {\rm cos} \Delta m t                                   \nonumber \\
       &&  - 2{\rm Re}\,\xi~{\rm sinh} \Delta\gamma t/2 
           - 2{\rm Im}\,\xi~{\rm sin}\Delta m t              ].
\label{rsdecay}
\end{eqnarray}
The time dependent probability for the  decay of a $\d0b$
to a final state $\overline f$,  
$\overline{P}_{\overline f}(t)$, may be obtained by making the substitutions
$\xi \rightarrow - \xi$ and
$F \rightarrow \overline{F}$ in the above equation. $F^*=\overline{F}$
is strictly true if CP (CPT) is not directly violated, which experimental
evidence suggests is very nearly true in charm decays.
$F=\langle f | T | D^0 \rangle$ represents the basic
transition amplitude for the decay $D^0\rightarrow f$,
$\Delta\gamma$ and $\Delta m$ are
the differences in physical decay widths and masses for the propagating eigenstates
and can be related to the usual mixing parameters~\cite{belleycp} $x=
\Delta M/\Gamma =-2\Delta m/\gamma$, $y = \Delta
\Gamma/{2\Gamma} =\Delta\gamma/\gamma$,
and $\gamma$ is
the sum of the physical decay widths.
Expressions for wrong-sign decay probabilities involve both CPT and T violation
parameters which only scale the probabilities, leaving the shape unchanged.
Using only right-sign decay modes, and assuming neglible direct CPT violation, the following asymmetry can be formed,
\begin{equation}
A_{\rm CPT}(t)  =  {{\overline{P}_{\overline f}(t) - P_{f}(t)}\over{\overline{P}_{\overline f}(t) + P_{f}(t)}},
\label{asym}
\end{equation}
which is sensitive to the CPT violating parameter $\xi$:
\begin{equation}
A_{\rm CPT}(t) = {{ 2{\rm Re}\,\xi~{\rm sinh}\Delta\gamma t/2 + 2{\rm Im}\,\xi~{\rm sin} \Delta m t} \over {
 ( 1 + |\xi|^2 ) {\rm cosh} \Delta\gamma t/2 + 
           (1 - |\xi|^2 ) {\rm cos} \Delta m t 
}}.
\label{acpt}
\end{equation}
Experiments show that $x,y$ mixing values are small~($<5\%$).
Equation~\ref{acpt}, for small $x,y$ and $t$, reduces to:
\begin{equation}
A_{\rm CPT}(t) = ({\rm Re}\,\xi~y-{\rm Im}\,\xi~x)~\Gamma~t.
\label{acpt_red}
\end{equation}

\mysection{Experimental and analysis details}

In this paper we search for a CPT violating signal
using data collected by the FOCUS Collaboration during an
approximately twelve month time period in 1996 and 1997 at Fermilab.  
FOCUS is an upgraded version of the E687 spectrometer. Charm particles
are produced by the interaction of high energy photons
(average energy $\approx$ 180 GeV for triggered charm states) 
with a segmented BeO target. In the target region, charged particles are
tracked by up to sixteen layers of microstrip detectors.  These detectors
provide excellent vertex resolution.  Charged particles
are further tracked by a system of five multi-wire proportional
chambers and are momentum analyzed by two oppositely 
polarized large aperture dipole magnets.  Particle
identification is accomplished by three multi-cell 
threshold ${\check{\rm C}}$erenkov detectors~
\cite{focuscerenkovnim}, two electromagnetic calorimeters,
an hadronic calorimeter and muon counters.

\begin{figure}
\begin{center}
\includegraphics[height=2.0in]{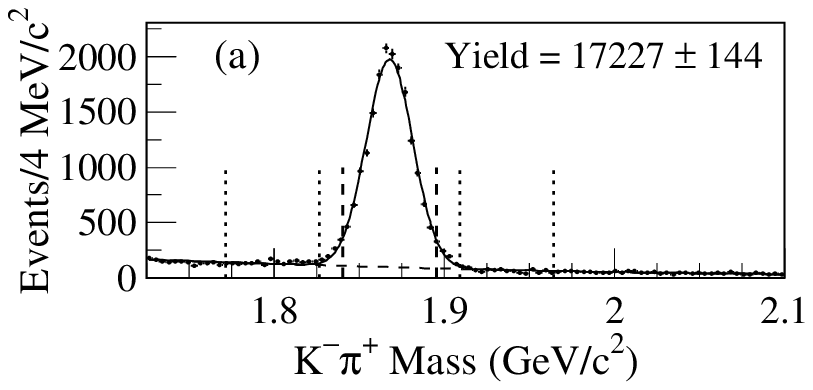}\\
\includegraphics[height=2.0in]{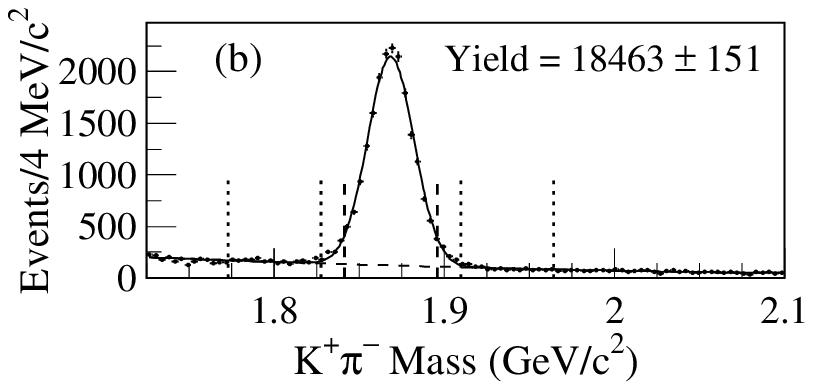}
\caption{Invariant mass of ($D^0\rightarrow K^-\pi^+$~(a); 
$\d0b\rightarrow K^+\pi^-$~(b)) for data (points) fitted with a
Gaussian signal and quadratic background (solid line). 
The vertical dashed lines indicate the signal
region, the vertical dotted lines indicate the sideband region.}  
\label{fig:dmass_split}
\end{center}
\end{figure}

We analyze the two right-sign hadronic 
decays $D^0 \rightarrow K^-\pi^+$ 
and $\d0b \rightarrow K^+\pi^-$.
We use the soft pion from the decay $D^{*+}\rightarrow D^0\pi^+$
to tag the flavor of the $D$ at production, and 
the kaon charge in the decay $D^0\rightarrow K^- \pi^+$
to tag the $D$ flavor at decay.
$D^0\rightarrow K^- \pi^+$ events are selected by requiring
a minimum detachment $\ell$ of the secondary (decay) vertex from the 
primary (production) vertex of 5 $\sigma_\ell$, where $\sigma_\ell$ is
the calculated uncertainty of the detachment measurement.  
The primary vertex is found using a candidate driven vertex
finder which nucleates tracks about a ``seed'' track 
constructed using the secondary vertex and the 
$D$ momentum vector. Both primary and secondary vertices
are required to have fit confidence levels greater
than 1\%.
The $D^*$-tag is implemented by requiring the
$D^*-D^0$ mass difference be within 3 MeV/c$^2$ of
the nominal value~\cite{pdg}.
A $\chi^2$-like variable called
$W_i \equiv $ --2~log(likelihood), where $i$ ranges
over electron, pion, kaon and proton hypotheses is used for particle
identification~\cite{focuscerenkovnim}. 
For the $K$ and the $\pi$ candidates we require $W_i$ to be no more
than four units greater than the smallest of the other three hypotheses
($W_i - W_{min} < 4$) which eliminates candidates that are
likely to be mis-identified. 
In addition, $D^0$ daughters must satisfy the slightly stronger
$K\pi$ separation criteria $W_\pi - W_K > 0.5$ for the $K$
and $W_K - W_\pi > -2 $ for the $\pi$.
Events in which the final state $K^-\pi^+$ is identified as
$\pi^-K^+$ and vice versa are removed by imposing a  hard
 $\check{\rm C}$erenkov cut on the sum of the two separations 
$((W_\pi - W_K)_K + (W_K - W_\pi)_\pi > 8) $. $K\pi$ pairs with highly 
asymmetrical momenta are more likely to be background than signal. A
cut is made on the momentum asymmetry,
$P_{A}=|({P_{K}-P_{\pi}})/({P_{K}+P_{\pi}})|$, to reject these
candidates. The best background rejection is achieved by applying
the cut in the following way, $P(D^0)>-160+280\times P_{A}$, where
$P(D^0), P_{K}$ and $P_{\pi}$ are the momenta of the $D$ and the daughter
kaon and pion respectively. 
To avoid large acceptance corrections due to the presence of a trigger
counter downstream of the silicon detector, we impose a fiducial cut on
the location of the primary vertex.
Figure~\ref{fig:dmass_split} shows the invariant mass distribution
for $D^*$-tagged, right-sign decays
$D^0\rightarrow K^-\pi^+$ and $\d0b\rightarrow K^+\pi^-$.
A fit to the mass distribution is carried out where a Gaussian
function for the signal and a second-order polynomial for the
background is used. The fit yields $17\,227 \pm 144$ $D^0$ and
 $18\,463 \pm 151$ $\d0b$ signal events. 

\begin{figure}[tbph!]
\begin{center}
\includegraphics[height=2.0in]{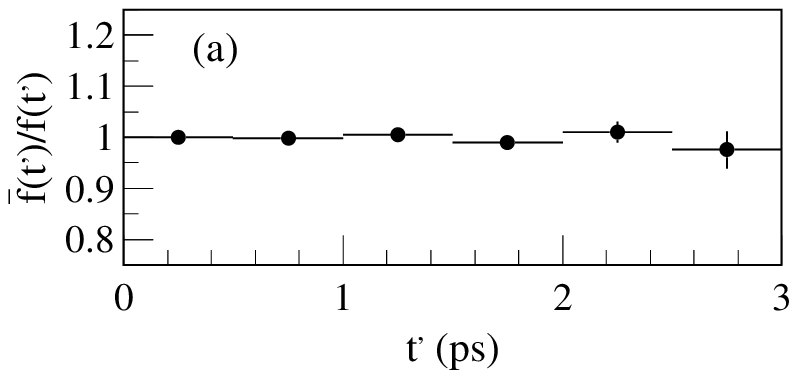}\\
\includegraphics[height=2.0in]{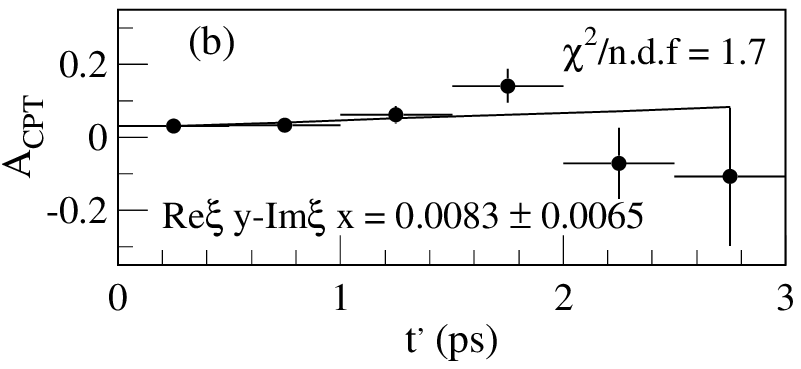}\\
\includegraphics[height=2.0in]{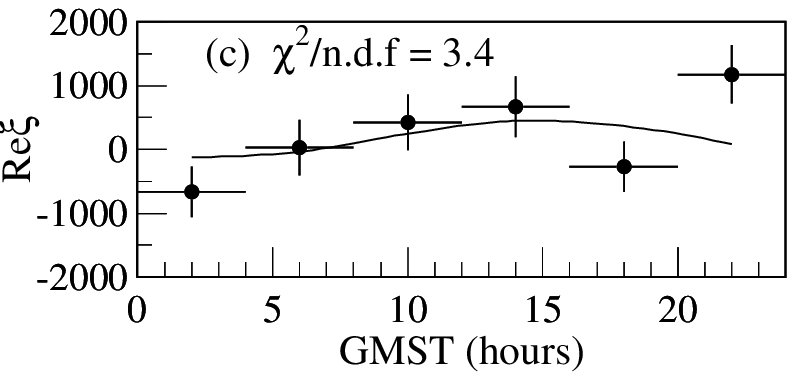}
\caption{(a) The ratio of the corrections; (b) $A_{\rm CPT}$ as a function of reduced proper time. The data
points represent the $A_{\rm CPT}$ as given in Equation~\ref{asym_exp} 
and the solid line represent the fit given in functional form by 
Equation~\ref{acpt_red}; (c) ${\rm Re}\,\xi$ as a function of Greenwich Mean
Sidereal Time~(GMST).}  
\label{fig:main_results}
\end{center}
\end{figure}

The proper time decay distribution is distorted by imposing 
a detachment cut between the primary and secondary vertices.
The reduced proper time,
defined as $t^{\prime}=(\ell - N\sigma_\ell)/(\beta\gamma c)$ where
$\ell$ is the distance between the primary and secondary
vertex, $\sigma_\ell$ is the resolution on $\ell$, and $N$
is the minimum detachment cut applied, removes this distortion. We
chose N=5 such that signal to background ratio was maximal. 
A simulation study was done measuring the differences in measured
values of $A_{CPT}$ and $\xi$ using $t^{\prime}$ in place of t in
Equation~\ref{asym_exp} and Equation~\ref{acpt_red}. 
The differences were found to be negligible compared to other
systematic uncertainties.
We plot the difference in right-sign events between $\d0b$ and
$D^0$ in bins of reduced proper time $t'$. The background subtracted
yields of right-sign $D^0$ and $\d0b$ were extracted by properly
weighting the signal region~($-2\sigma, +2\sigma$), the low mass
sideband~($-7\sigma, -3\sigma$) and high mass sideband~($+3\sigma,
+7\sigma$), where $\sigma$ is the width of the Gaussian. 
For each data point, these yields were used in forming the ratio:
\begin{equation}
A_{\rm CPT}(t') = {{\overline{Y}(t')-Y(t'){{\overline{f}(t')}\over{f(t')}}}\over{
\overline{Y}(t')+Y(t'){{\overline{f}(t')}\over{f(t')}}}},
\label{asym_exp}
\end{equation}
where $\overline{Y}(t')$ and $Y(t')$ are the yields for
 $\d0b$ and $D^{0}$ and $\overline{f}(t')$, $f(t')$ are
their respective correction functions. The functions 
$\overline{f}(t')$ and $f(t')$
account for geometrical acceptance, detector and reconstruction
efficiencies, and absorption of parent and daughter
particles in the nuclear matter of the target.
The correction functions are determined using a detailed Monte
Carlo~(MC) simulation using \textsc{PYTHIA}~\cite{pythia61}.
The fragmentation is done using the Bowler modified Lund string model. 
\textsc{PYTHIA} was tuned  using many production 
parameters to match various data production variables such as charm 
momentum and primary multiplicity.
The shapes of the $f(t^{\prime})$ and $\overline{f}(t^{\prime})$
functions are obtained
by dividing the reconstructed MC $t^{\prime}$ distribution by a pure
exponential with the MC generated lifetime.
The ratio of the correction functions, shown in Figure~\ref{fig:main_results}(a),
enters explicitly in Equation~\ref{asym_exp} and its effects on the
asymmetry are less than 1.3\% compared to when no corrections are applied. 
The FOCUS data contains more $\d0b$ than $D^0$ decays due to
production asymmetry~\cite{e687}. 
The effect on the $A_{\rm CPT}$ distribution is to add a constant 
offset, which is accounted for in the fit. 

\mysection{Fitting for the Assymmetry}

The $A_{\rm CPT}$ data in Figure~\ref{fig:main_results}(b) are fit to a 
line using the form of Equation~\ref{acpt_red} plus a constant offset. 
The value of $\Gamma$ used in the fit is $\Gamma=1.6\times10^{-12}$~GeV.
The result of the fit is ${\rm Re}\,\xi~y-{\rm Im}\,\xi~x=0.0083\pm0.0065$.

\mysection{Lorentz Violation}

Any CPT and Lorentz violation within the Standard Model 
is described by the Standard Model Extension (SME) 
proposed by Kosteleck\'y at al.~\cite{colladay-kostelecky}.
In quantum field theory,
the CPT violating parameter $\xi$ must generically depend on lab momentum,
spatial orientation, 
and sidereal time~\cite{kostelecky-prl,kostelecky}.
The SME can be used to show that Lorentz violation in the $D$ system
is controlled by the four vector $\Delta a_{\mu}$. 
The precession of the experiment with the earth relative to
the spatial vector $\vec{\Delta a}$ 
modulates the signal for CPT violation, thus
making it possible to separate the components of $\Delta a_{\mu}$. 
The coefficients for Lorentz violation depend on the flavor of the 
valence quark states and are model independent. 
In the case of FOCUS, where $D^0$ mesons in the lab frame
are highly collimated in the forward direction and under the
assumption that $D^0$ mesons are uncorrelated, the $\xi$ parameter assumes the 
following form~\cite{kostelecky}:
\begin{eqnarray}
\xi(\hat{t}, p)  & = & {\gamma(p)\over{\Delta\lambda}}[ \Delta a_0 + \beta \Delta a_Z {\rm cos} \chi \nonumber \\
        & &
+ \beta {\rm sin} \chi (\Delta a_Y {\rm sin} \Omega \hat{t} + \Delta a_X {\rm cos} \Omega \hat{t}) ].
\label{xi}
\end{eqnarray}

$\Omega$ and $\hat{t}$ are the sidereal frequency and time respectively, 
$X, Y, Z$ are non-rotating coordinates with $Z$ aligned along the
Earth's rotation axis, $\Delta\lambda=\Gamma(x-iy)$, and 
$\gamma(p) = \sqrt{1 + P^2_{D^0}/m_{D^0}^2}$. Binning in sidereal time 
$\hat{t}$ is very useful because it 
provides sensitivity to components $\Delta a_X$ and $\Delta a_Y$.    
Since Equation~15 of Reference~\cite{kostelecky} translates into ${\rm Re}\,\xi\,y - {\rm Im}\,\xi\,x = 0$,
setting limits on the coefficients of Lorentz violation requires expanding
the asymmetry in Equation~\ref{acpt} to higher (non-vanishing) terms. 
In addition,
the interference term of right-sign decays with the doubly Cabibbo
suppressed (DCS) decays must also be included since it gives a comparable
contribution. One can follow the procedure given by equations
[16] to [20] of Reference~\cite{kostelecky} where the basic transition
amplitudes $<f|T|\overline{P^0}>$ and $<\overline{f}|T|P^0>$ are not
zero but are DCS amplitudes. After Taylor expansion the asymmetry can be
written as:
\begin{eqnarray}
A_{\rm CPT} &=& \frac{{\rm Re}\,\xi (x^2 + y^2) (t/\tau)^2}{2x}
\nonumber \\
             & &
\left[ \frac{xy}{3} (t/\tau) + \sqrt{R_{\rm DCS}}\left(
x\,\cos{\delta}+ y\,\sin{\delta} \right) \right],
\label{new_asym}
\end{eqnarray}
where $R_{\rm DCS}$ is the branching ratio of DCS relative to
right-sign decays and $\delta$ is the strong phase between the DCS
and right-sign amplitudes.

\mysection{Fitting for LV Parameters}

We searched for a sidereal time dependence~\footnote{Sidereal time is
a time measure of the rotation of the Earth with respect to the stars, rather than the Sun. Sidereal day is shorter than the normal solar day by about 4 minutes.} by dividing our data sample into
four-hour bins in Greenwich Mean Sidereal 
Time (GMST)~\cite{JeanMeeus}, where for each bin we repeated
our fit in $t^\prime$ using the asymmetry given by
Equation~\ref{new_asym} and extracted ${\rm Re}\,\xi$. The value of
$\sqrt{R_{\rm DCS}}$ used in the fit is taken from
Reference~\cite{pdg} and it is $0.06$. 
The resulting distribution, shown in Figure~\ref{fig:main_results}(c),
was fit using
Equation~\ref{xi} and the results for the expressions involving 
coefficients of Lorentz violation in the SME were $C_{0Z}\equiv N(x,y,\delta)(\Delta a_0 + 0.6\,\Delta a_Z)
=(1.0\pm1.1)\times 10^{-16}$~GeV, $C_{X}\equiv N(x,y,\delta)\Delta a_X=(-1.6\pm2.0)\times
10^{-16}$~GeV, and $C_{Y}\equiv N(x,y,\delta)\Delta a_Y=(-1.6\pm2.0)\times 10^{-16}$~GeV, where
$N(x,y,\delta)=[{xy}/3 + 0.06\,(
x\,\cos{\delta}+ y\,\sin{\delta})]$ is the factor which carries the
$x,y$ and $\delta$ dependence. 
The angle between the FOCUS spectrometer axis and the Earth's
rotation axis is approximately $\chi = 53^\circ ({\rm cos}{\chi} =
0.6)$. We average over all $D^0$ momentum so $\langle \gamma(p) \rangle
\approx \gamma(\langle p \rangle) = 39$ and $\beta = 1$. We also
touched base with the previous measurements
for the kaon $r_K$ and B meson $r_B$ by constructing a similar
quantity $r_D$~\cite{kostelecky99},
$r_D=|\Delta\Lambda|/m_{D^0}=\beta^\mu\Delta a_{\mu}/m_{D^0}=|\overline{\xi}||\Delta\lambda|=\gamma(p)|\Delta
a_0 + 0.6\,\Delta a_Z|/m_{D^0}$. The result for $N(x,y,\delta)\,r_D$ is:
$N(x,y,\delta)\,r_D=(2.3\pm2.3)\times 10^{-16}$~GeV. Although it may
seem natural to report $r_D$, the parameter $r_D$ (and $r_K$, $r_B$) has
a serious defect: in quantum field theory, its value changes 
with the experiment. This is because it is a combination of the 
parameters $\Delta a_{\mu}$ with coefficients controlled by the $D^0$ meson 
energy and direction of motion. 
The sensitivity would have been best if $\chi = 90^\circ$.

\section{Systematic Errors}

Previous analyses have shown that MC absorption corrections are very small
~\cite{focusycp}.
The interactions of pions and
kaons with matter have been measured but no equivalent data exists for
charm particles. To check any systematic effects associated
with the fact that the charm particle cross section is unmeasured, 
we examined several variations of $D^0$ and $\d0b$ cross sections.
The standard deviation of these
variations returns systematic uncertainties of $\pm0.0017$, 
$\pm0.3\times 10^{-16}$~GeV, $\pm0.0\times 10^{-16}$~GeV, and
$\pm0.1\times 10^{-16}$~GeV to our measurements of 
${\rm Re}\,\xi~y-{\rm Im}\,\xi~x$, $C_{0Z}$, $C_X$, and $C_Y$
respectively. 

In a manner similar to the S-factor method used by the Particle Data
group PDG~\cite{pdg} we made eight statistically independent
samples of our data in order to look for systematic effects. We split
the data in four momentum ranges and two years. The split in year was
done to look for effects associated with target
geometry and reconstruction due to the addition of four silicon planes
near the targets in January, 1997~\cite{targetsilicon}. 
We found no contribution to our measurements of ${\rm Re}\,\xi~y-{\rm
Im}\,\xi~x$ and $C_{0Z}$. The contributions for
$C_X$ and $C_Y$ were $\pm1.3\times 10^{-16}$~GeV and $\pm1.6\times
10^{-16}$~GeV respectively.  
 
We also varied the bin widths and the position of the sidebands to
assess the validity of the background subtraction method and the
stability of the fits. The standard deviation of these
variations returns systematic uncertainties of $\pm0.0012$, $\pm0.3\times
10^{-16}$~GeV, $\pm0.9\times 10^{-16}$~GeV, and $\pm0.5\times
10^{-16}$~GeV to our measurements of 
${\rm Re}\,\xi~y-{\rm Im}\,\xi~x$, $C_{0Z}$, $C_X$, and $C_Y$ respectively.

Finally, to uncover any unexpected systematic uncertainty, we varied
our $\ell/\sigma_\ell$ and $W_\pi - W_K$ requirements and
the standard deviation of these variations returns systematic uncertainties of
$\pm0.0036$, $\pm1.5\times 10^{-16}$~GeV, $\pm1.0\times 10^{-16}$~GeV,
and $\pm1.1\times 10^{-16}$~GeV to our measurements of 
${\rm Re}\,\xi~y-{\rm Im}\,\xi~x$, $C_{0Z}$, $C_X$, and $C_Y$ respectively.

Contributions to the systematic uncertainty are summarized in 
Table~\ref{tb_syst1} and Table~\ref{tb_syst2}. 
Taking contributions to be uncorrelated 
we obtain a total systematic uncertainty
of $\pm 0.0041$ for ${\rm Re}\,\xi\,y - {\rm Im}\,\xi\,x$, 
$\pm1.6\times10^{-16}$~GeV for $C_{0Z}$, 
$\pm1.9\times10^{-16}$~GeV for $C_X$, and
$\pm2.0\times10^{-16}$~GeV for $C_Y$.
\begin{table}[htp]
\caption{\label{tb_syst1}Contributions to the systematic uncertainty.}
\begin{center}
\begin{tabular}{l|c|c}
Contribution & ${\rm Re}\,\xi\,y - {\rm Im}\,\xi\,x$ & $C_X$~(GeV) \\
\hline
Absorption  & $\pm 0.0017$ & $\pm0.0\times 10^{-16}$\\
Split sample & $\pm 0.0000$ & $\pm1.3\times 10^{-16}$ \\
Fit variant  & $\pm 0.0012$ & $\pm0.9\times 10^{-16}$ \\
Cut variant  & $\pm 0.0036$ & $\pm1.0\times 10^{-16}$\\
\hline
Total       & $\pm 0.0041$ & $\pm1.9\times 10^{-16}$\\
\end{tabular}
\end{center}
\end{table}

\begin{table}[h]
\caption{\label{tb_syst2}Contributions to the systematic uncertainty.}
\begin{center}
\begin{tabular}{l|c|c}
Contribution &$C_{0Z}$~(GeV) &  $C_Y$~(GeV)\\
\hline
Absorption  & $\pm 0.3\times 10^{-16}$ & $\pm0.1\times 10^{-16}$\\
Split sample & $\pm 0.0\times 10^{-16}$ & $\pm1.6\times 10^{-16}$ \\
Fit variant  & $\pm 0.3\times 10^{-16}$ & $\pm0.5\times 10^{-16}$ \\
Cut variant  & $\pm 1.5\times 10^{-16}$ & $\pm1.1\times 10^{-16}$\\
\hline
Total       & $\pm 1.6\times 10^{-16}$ & $\pm2.0\times 10^{-16}$\\
\end{tabular}
\end{center}
\end{table}
 
\mysection{Summary}

We have performed the first search for CPT and Lorentz violation in 
neutral charm meson oscillations. We have measured 
${\rm Re}\,\xi\,y - {\rm Im}\,\xi\,x = 0.0083 \pm 0.0065 \pm 0.0041$
which lead to a 95\% confidence level limit of
$-0.0068 < {\rm Re}\,\xi\,y - {\rm Im}\,\xi\,x < 0.0234$.
As a specific example, assuming $x=0$ or ${\rm Im}\,\xi = 0$ and $y = 1\%$,
one finds ${\rm Re}\,\xi = 0.83 \pm 0.65 \pm 0.41$ with a 95\%
confidence level limit of $-0.68 < {\rm Re}\,\xi < 2.34$.  Within the 
Standard Model Extension, we set three independent first limits on the
expressions involving coefficients of Lorentz violation of
 $(-2.8<N(x,y,\delta)(\Delta a_0 + 0.6\,\Delta
a_Z)<4.8)\times 10^{-16}$~GeV, $(-7.0<N(x,y,\delta)\Delta
a_X<3.8)\times 10^{-16}$~GeV, and $(-7.0<N(x,y,\delta)\Delta a_Y<3.8)\times
10^{-16}$~GeV. As a specific example, assuming $x=1\%$, $y = 1\%$ and
$\delta=15^\circ$
one finds the 95\% limits on the coefficients of Lorentz violation of $(-3.7<\Delta a_0 + 0.6\,\Delta
a_Z<6.5)\times 10^{-13}$~GeV, $(-9.4<\Delta a_X<5.0)\times 10^{-13}$~GeV,
and $(-9.3<\Delta a_Y<5.1)\times 10^{-13}$~GeV. The measured values
are consistent with no CPT or Lorentz invariance violation.

\mysection{Acknowlegments}
We wish to acknowledge the assistance of the staffs of Fermi National
Accelerator Laboratory, the INFN of Italy, and the physics departments
of the collaborating institutions. This research was supported in part by the
U.~S.
National Science Foundation, the U.~S. Department of Energy, the Italian
Istituto Nazionale di Fisica Nucleare and 
Ministero della Istruzione Universit\`a e
Ricerca, the Brazilian Conselho Nacional de
Desenvolvimento Cient\'{\i}fico e Tecnol\'ogico, CONACyT-M\'exico, and
the Korea Research Foundation of the  
Korean Ministry of Education.


\begin{thebibliography}{10}
\bibitem{kostelecky-95}
V.~A.~Kosteleck\'y and R.~Potting, \PRD~51, 3923 (1995).
\bibitem{xing} Z.~Xing, \PRD~55, 196 (1997).
\bibitem{kostelecky} V.~A.~Kosteleck\'y, \PRD~64, 076001 (2001).
\bibitem{alanbook}
V.~A.~Kosteleck\'y,
 {\it CPT and Lorentz Symmetry II
}; Singapore, World Scientific Publishing Co.Pte.Ltd. (2002).
\bibitem{lavoura} L.~Lavoura, Ann.~Phys.~207, 428 (1991).
\bibitem{kostelecky99}
V.~A.~Kosteleck\'y, \PRD~61, 016002 (1999), 
section 2 of the paper gives
relationships between formalisms.
\bibitem{ktev}
Y.~B.~Hsiung,
  for the KTeV Collaboration,
Proc.\ QCD Euroconference 99, Montpellier, France, July 7-13, 1999.  Also
available as FERMILAB-Conf-99/338-E (KTeV).
\bibitem{ktevlv} H.~Nguyen,
  CPT results from KTeV, hep-ex/0112046.
\bibitem{opal}
OPAL Collab., K.~Ackerstaff \etal, Z.Phys.C~76, 401 (1997).
\bibitem{belle}
Belle Collab., K.~Abe \etal,
  Phys. Rev. Lett.~86, 3228 (2001).
\bibitem{focusycp}
FOCUS Collab., J.~M.~Link \etal, \PLB~485, 62 (2000).
\bibitem{belleycp}
Belle Collab., K.~Abe \etal, Phys. Rev. Lett.~88, 162001 (2002).
\bibitem{babary}
B.~Aubert \etal,
  Search for a Lifetime Difference in $D^0$ Decays, hep-ex/0109008.
\bibitem{cleo}
R.~Godang \etal, \PRL~84, 5038 (2000).
\bibitem{link}
J.~M.~Link \etal, Phys. Rev. Lett.~86, 2955 (2001).
\bibitem{focuscerenkovnim} J.~M.~Link \etal,
 Nucl.\ Instrum.\ Methods\ Phys.\ Res.,\ Sect.\ A~484, 270 (2002).
\bibitem{pdg}
D.~E.~Groom \etal, \EPJ~15, 1 (2000).
\bibitem{pythia61}
T.~Sj\"{o}strand \etal, Comput.\ Phys.\ Commun.~135, 238 (2001).
\bibitem{e687}
P.~L.~Frabetti \etal, \PLB~370, 222 (1996).
\bibitem{colladay-kostelecky}
D.~Colladay and V.~A.~Kosteleck\'y, \PRD~55, 6760 (1997); \PRD~58, 116002 
(1998).
\bibitem{kostelecky-prl}
V.~A.~Kosteleck\'y, \PRL~80, 1818 (1998). 
\bibitem{JeanMeeus}
J.~Meeus, {\it Astronomical
Algorithms}; Richmond (Virginia), Willmann-Bell Inc. (1991); Chap.11.
\bibitem{targetsilicon}
J.~M.~Link et al.,
  FERMILAB-Pub-02/069-E, hep-ex/0204023, to be published in Nucl.\
  Instrum.\ Methods\ Phys.\ Res.,\ Sect.\ A.

\end{thebibliography}
\end{document}